# Discovery of Li$_2$(Pd,Pt)$_3$B superconductors

P. Badica[a,b,*], K. Togano[c], H. Takeya[c], K. Hirata[c], S. Awaji[a] and K. Watanabe[a]

[a]*High Field Laboratory for Superconducting Materials, Institute for Materials Research, Tohoku University, Sendai, 980-8577, Japan*

[b]*National Institute of Materials Physics, Bucharest-Magurele, 077125, Romania*

[c]*National Institute for Materials Science, 1-2-1 Sengen, Tsukuba, 305-0047, Japan*

Critical temperature $T_c$ of the Li$_2$(Pd$_{1-x}$Pt$_x$)$_3$B was reported to be 7-8K for x=0 and 2.2-2.8K for x=1. In this article we present our preliminary results on behavior of magnetization-temperature curves with starting composition of Pd-B precursor, y-Li concentration in Li$_y$Pd$_3$B and post-annealing of the Pd-end compound. Results suggest that to maximize $T_c$ ratio Pd:B should be close to 3:1, while y-Li has to be optimum. The lowest $T_c$ for Li$_y$Pd$_3$B was 4.4-4.6K, while post-annealings at 560°C allowed enhancement of $T_c$ up to 8.2-8.4K. Compositions Li$_2$Z$_3$B with Z=Ni, Ru, Rh, Re, Ag are not superconducting down to 1.8K. Exception is composition with Re showing superconductivity due to Re$_3$B compound. All samples were prepared by arc melting.

## 1. Introduction

Superconductivity was recently found in Li$_2$(Pd$_{1-x}$,Pt$_x$)$_3$B solid solution [1-3], and it was observed in the entire x=0-1 range.

Material has an interesting structure [4, 5] that can be described as cubic, of antiperovskite-like type. It is composed of Pd$_6$B distorted octahedrons which resemble octahedrons from the structure of high-$T_c$ superconductors (HTS), ruthenates, cobaltates and MgNi$_3$C. Distorsion was suggested to influence superconductivity [6] through strong electronic correlations related to three kinds of Z-Z bond length, with Z=Pd or Pt. On the other hand, PES and NMR experimental results from refs [7, 8] indicate that correlation effects play a negligible role for the physical properties of the superconductor.

Although electronic structure calculations [9-11] based on a new interpretation of the crystal structure led to similar conclusions as those resulted from the experimental works [7, 8], there are several unclear aspects in the complex electronic structure of these materials which point toward significant differences between Pd- and Pt-end compounds. It also suggests possibility of special or unconventional effects which, for example, might determine $T_c$ of the Pt-end compound considered unusually low.

The structure lacks mirror-like symmetry [10, 11] as for CePt$_3$Si [12] or UIr [13]. From this and other points of view, Pt- and Pd- end compounds are very different from MgNi$_3$C [11]. It was anticipated that Li$_2$(Pd$_{1-x}$,Pt$_x$)$_3$B might be an ideal candidate to study the influence of this type of broken symmetry on superconductivity.

To explain low $T_c$ in Pt-end superconductor, in ref [14] authors considered that broken inversion symmetry might connect with observed differences in the behavior of the penetration depth λ and proposed that both compounds show unconventional superconductivity. Namely, Li$_2$Pd$_3$B behaves as a double-gap superconductor similar to MgB$_2$, while Li$_2$Pt$_3$B has nodes in the energy gap. The result contradicts most of the mentioned studies, refs [15, 16] and penetration depth data obtained from μSR experiments [17], all of them suggesting that at least Li$_2$Pd$_3$B is an *s*-wave BCS superconductor with only one isotropic energy gap.

Gradual introduction of Ni [18] in $Li_2(Pd_{1-x}Ni_x)B$ with x=0-0.2 and application of external pressure (0-3GPa) on $Li_2Pd_3B$ decrease $T_c$. But, the rate of $T_c$-decrease ($dT_c/da$, $a$=lattice parameter) with Ni substitution was found to be lower than in the case of Pt-substitution. Authors also wrote that pressure experiments indicate that $Li_2Pd_3B$ is probably a conventional superconductor. No conclusion was made about superconductivity in $Li_2(Pd_{1-x}Ni_x)_3B$.

Material $Li_2(Pd_{1-x},Pt_x)_3B$, x=0-1, is the first boride superconductor containing alkaline element Li and late transition elements Pd and/or Pt. No binary or ternary Pd containing boride superconductors were reported, although Pd gives the highest $T_c$ in $RE(TM)_2B_2C$ quaternary boride system with RE=rare earth metal and TM=Ni, Pt or Pd [19].

Presented features and encountered fundamental details, some of them under debate, make investigation of this material of high interest and this is promising for further developments with possibly high impact on understanding of superconductivity. To have a reliable progress in this direction high quality samples are required. At present moment, Li-concentration control in the samples is very difficult and hence $T_c$ is showing large scattering and transition width is relatively broad. Related questions are: what is the maximum $T_c$ and is nonstoichiometry necessary in this regard as for other boride superconductors [20]?

To answer these questions, we have decided to prepare and investigate by magnetization-temperature (M(T)) measurements stoichiometric and nonstoichiometric samples of $Li_2Pd_3B$, and we applied post-annealing heat treatments for samples with Pd:B ratio equal to 3:1. We also present several attempts to synthesize other $Li_2Z_3B$ (Z=metal) materials and to test their superconductivity.

### 3. Experimental

Samples of Li-Pd-B (denoted LPB) were prepared by arc melting [1] in Ar atmosphere (1atm). We applied a two-step melting procedure. In the first one, Pd-B alloys with different compositions (Pd:B = 2:1, 2.5:1, 3:1 and 5:1) were prepared. Weight loss was less than 0.3%. In the second step, as-prepared PdB-alloy (about 200mg) was placed on a freshly cut Li-metal plate (10-50mg) and arc re-melted for few seconds ($\leq$ 10s). Intake of Li into PdB alloy is very fast. Concentration of Li in the samples was roughly estimated from the weight gain of PdB alloy. Precise control of Li concentration and homogeneity is very difficult due to its high volatility (see *4.2.* and *4.3.*).

Same synthesis method was used to prepare samples with starting composition $Li_2Z_3B$ where Z= Ni, Ru, Rh. Re and Ag (denoted 231Z).

Samples (as-prepared buttons) of stoichiometric 231Pd were broken into smaller pieces. Usually the shape of these pieces is of rectangular plate suggesting that they are single crystal-like. Several such crystals were selected and post-annealed in a tube furnace with Ar-atmosphere. Annealing conditions were 530 and 560°C for 4 and 0.5h, respectively. Furnace cooling was applied.

Samples were characterized by magnetization measurements using a SQUID magnetometer (Quantum Design). X-ray diffraction (XRD) was performed with a PANalytical/Phillips diffractometer (CuK$\alpha$ radiation).

### 4. Results and Discussion

*4.1. Samples Li-Pd-B with different Pd:B ratio*

PdB-alloys do not show superconductivity. On the other hand Li-Pd-B samples are superconducting. Magnetization M(T) curves for samples prepared from PdB alloys with different Pd:B ratio are presented in Fig. 1. These curves are for our LPB samples with the highest onset critical temperature, $T_c$. Remarkably, all curves are showing approximately the same $T_c$ of about 7.8-8K and just transition width and magnitude of the magnetization signal at 2K are systematically decreasing and increasing from sample A to C, respectively. Such behavior indicates that the grains of the highest quality have the same $T_c$ and these grains have the ratio close to Pd:B=3:1 as in the $Li_2Pd_3B$. Variation of $T_c$ with Li concentration in y31 samples is presented in *4.2*. XRD data (Fig. 2) taken on A-C ground powder samples are supporting above conclusion: better superconducting M(T) transition in the sample with Pd:B=3:1 is accompanied by the increase of the 231Pd phase concentration, while the impurity phases level decreases (from A to C). For reference, we have also included in Fig. 2 the XRD pattern of the starting Pd:B=3:1 (031 sample) alloy.

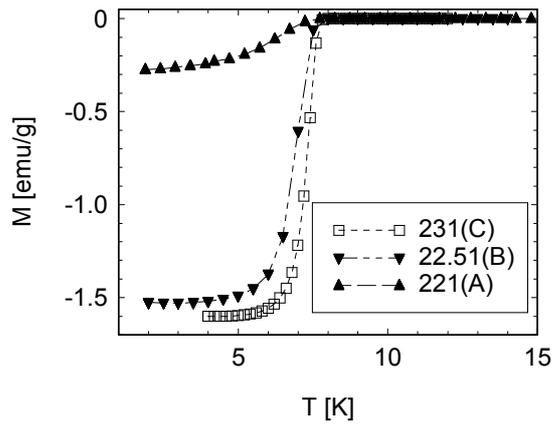

Fig. 1 Zero-field-cooled magnetization vs. temperature for LPB with different Pd:B ratio measured for a magnetic field H of 100Oe.

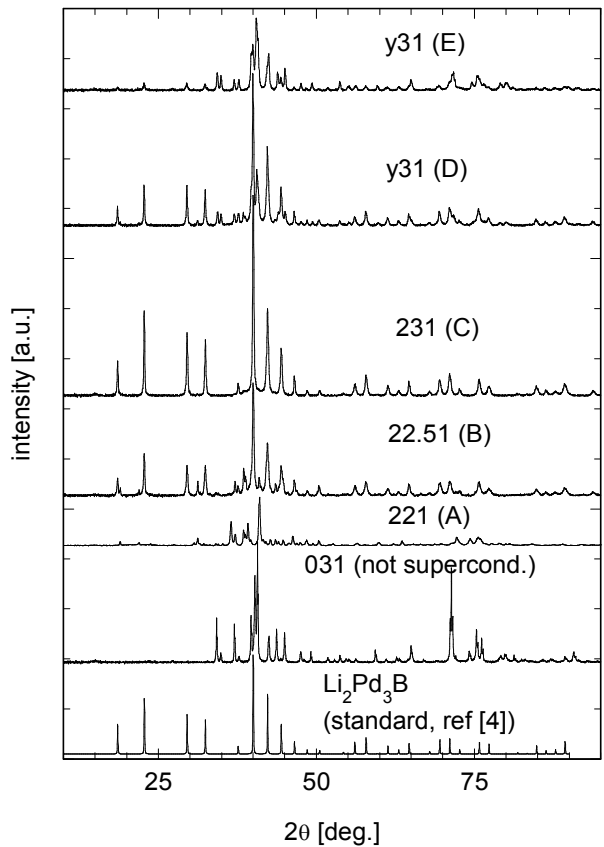

Fig. 2 XRD powder patterns of the LPB samples from Fig. 1 and $Li_yPd_3B$ (y31) samples from *4.2*. Standard pattern [4] of stoichiometric 231Pd is also shown.

Roughly, data suggest that to obtain superconductivity and to maximize $T_c$, Pd:B non-stoichiometry is likely not required. It also proves once more that 231Pd phase is superconducting.

*4.2. Samples $Li_yPd_3B$ (y31)*

Samples y31 prepared from the alloy Pd:B=3:1 may have a different $T_c$ than 7.8-8K (Fig. 3). XRD patterns (Fig. 2) show that these samples contain 231Pd phase. Impurity phases were also detected as for the samples presented in the section *4.1*. However, situation is very different from the one described in *4.1.*: the approximately parallel shift in the M(T) curve is suggesting that in these samples the amount of Li in the 231 phase is different and deficiency of Li is resulting in significantly lower $T_c$. Samples with y>2 usually had $T_c$=7.8-8K and a broader transition (Fig. 3). We note that ICP composition of a sample ($T_c$=7.8K) similar to the sample 231(C) was 1.9:2.97:1, i.e. close to ideal 231Pd composition and in agreement with y estimated from weight data. The lowest $T_c$ for our y31 samples was 4.4-4.6K. Possibility that one of the impurity phases is responsible for superconductivity at 4.4-4.6K cannot be totally excluded, but our results suggest that the probability is rather low.

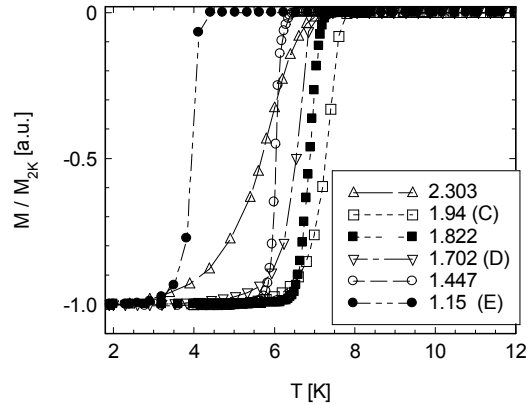

Fig. 3 Reduced zero-field-cooling (ZFC) magnetization curves (H=100Oe) of (y31)Pd samples. Values of y estimated from weight data are given in the legend. Samples C, D and E are the same from Fig. 2.

Data of weight intake as well as of ICP show average composition of the sample. For a more precise analysis of magnetization-composition dependency, knowledge of the local stoichiometry of the superconducting grains from each sample is necessary.

Behavior of M(T) curves from *4.1.* and *4.2.* can be probably understood within the proposed description of superconductivity for 231Pd(Pt) in which Li contribution to electron-phonon coupling is negligible, and its almost entire valence charge is transferred to B-Pd(Pt) network [9-11]. Li seems to have the role of charge reservoir controlling the value of $T_c$, but more research is necessary. One related unsolved point of interest is how the structure of 231Pd can accommodate Li deficiency or excess.

*4.3. Annealing of $Li_2Pd_3B$ samples*

In ref. [18] authors found a $T_c$ (onset) of 8.2K from *ac* susceptibility measurements employing a laboratory made setup. Samples were prepared by a combined arc melting and solid state reaction method. On the other hand, Takeya et al [21] prepared samples by arc melting and solid state reaction using rf and conventional heating. Samples were measured by a SQUID magnetometer. The maximum $T_c$ (onset) from the M(T) was not exceeding 8K and the highest value was obtained for the arc melted sample, although this sample was not showing the sharpest transition. Observed differences might be due to non-homogeneity as a consequence of fine stoichiometry variations and/or residual stress. In these circumstances we have considered application of post-annealing treatments to 231Pd superconductor. Before presenting results we emphasize that all samples in this article were prepared and measured by the same equipment.

One as-prepared sample was annealed at 530°C for 4h in Ar (Fig. 4). Annealed sample has shown same $T_c$ (7.6-7.8K) as the parent one, but the magnitude of both FC and ZFC magnetization curves enhances, suggesting the increase of the superconducting volume fraction. At the same time transition width become somewhat broader.

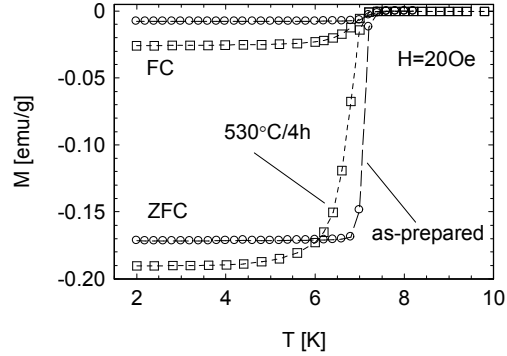

Fig. 4 Magnetization curves for as-prepared and annealed sample.

Considering this result, the next two samples were annealed for shorter time, 30min. Temperature was 560°C and we used crystals of regular shape selected from one ground as-prepared button. Reduced magnetization curves for the crystals before (K1, L1) and after annealing (K2, L2, respectively) are shown in Fig. 5.

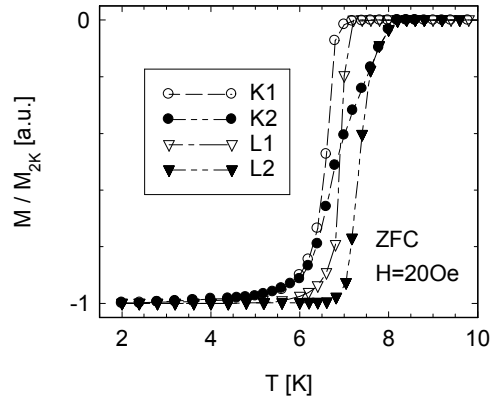

Fig. 5 Reduced magnetization curves for 213Pd crystals before and after post-annealing in Ar at 560°C for 30min.

Transition width is broader after annealing, and the effect is strong for (K) and is almost imperceptible for (L) crystal. $T_c$ has a clear tendency for enhancement. Both annealed crystals are showing weak diamagnetic signals at 8.2K. We conclude that $T_c$ of the annealed samples is 8.2-8.4K. The result is not surprising considering the $T_c$ of 8.2K from ref. [18] and the offset-$T_c$ of 8.2K reported from the resistivity measurements [1, 3, 21]. Recalling our results from *4.1.* and *4.2.*, it seems that post-annealing allows to bring the stoichiometry, especially that of Li, at its optimum for which $T_c$ is maximum. However, the effect is missing its clear explanation and further studies are necessary to understand it.

Another issue of interest is that crystals (K1) and (L1) extracted from the as-prepared button show differences in the reduced magnetization curves. Notable are the different width and inclination of the superconducting transition. Hence, homogeneity of the as-prepared button samples is not perfect and care should be taken when using such samples in different experiments. It is recommended to use small crystals and to improve their quality through optimized post-annealing treatments.

*4.4. Samples $Li_2Z_3B$, Z=Ni, Ru, Rh, Re, Ag*

Samples of $Li_2Z_3B$ where Z= Ni, Ru, Rh, Re and Ag did not show superconductivity down to 1.8K. Exception is 231Re. $T_c$ was 4.6K. A sample 031Re had the same pattern in M(T) curve, and $T_c$ was the same. Therefore, superconductivity in 231Re is due to phase $Re_3B$ for which it was reported in ref. [22] a $T_c$=4.65K. XRD data show that samples 231Z consist mainly of binary boride phases such as $Ni_3B$, $Ru_7B_3$, $Rh_7B_3$, $Re_3B$ (some $ReB_2$), and elemental Li, B and Ag. For the sample 231Ag the most intensive peaks were not identified suggesting formation of an un-known ternary phase. Some AgLi was also detected.